\documentclass[%
 showpacs,
 amsmath,amssymb,
]{revtex4-1}

\usepackage{graphicx}
\usepackage{dcolumn}
\usepackage{bm}

\usepackage{CJK}

\usepackage{color}
\usepackage{amssymb}
\usepackage{amsfonts}

\usepackage[colorlinks,urlcolor=blue,linkcolor=blue,citecolor=blue,anchorcolor=blue]{hyperref}

\begin{document}

\preprint{APS/123-QED}

\title{Transport properties in the photonic super-honeycomb lattice --- a hybrid fermionic and bosonic system}

\author{Hua Zhong$^1$}
\author{Yiqi Zhang$^{1,2}$}
\email{zhangyiqi@mail.xjtu.edu.cn}
\author{Yi Zhu$^3$}
\author{Da Zhang$^1$}
\author{Changbiao Li$^{1,2}$}
\author{Yanpeng Zhang$^{1}$}
\author{Fuli Li$^2$}
\author{Milivoj R. Beli\'c$^4$}
\author{Min Xiao$^{5,6}$}
\affiliation{%
$^1$Key Laboratory for Physical Electronics and Devices of the Ministry of Education \& Shaanxi Key Lab of Information Photonic Technique,
Xi'an Jiaotong University, Xi'an 710049, China \\
$^2$Department of Applied Physics, School of Science, Xi'an Jiaotong University, Xi'an 710049, China\\
$^3$Zhou Pei-Yuan Center for Applied Mathematics, Tsinghua University, Beijing 100084, China \\
$^4$Science Program, Texas A\&M University at Qatar, P.O. Box 23874 Doha, Qatar \\
$^5$Department of Physics, University of Arkansas, Fayetteville, Arkansas 72701, USA \\
$^6$National Laboratory of Solid State Microstructures and School of Physics, Nanjing University, Nanjing 210093, China
}%

\date{\today}

\begin{abstract}
\noindent
  We report on transport properties of the \textit{super-honeycomb lattice},
  the band structure of which possesses a flat band and Dirac cones, according to the tight-binding approximation.
  This super-honeycomb model combines the honeycomb lattice and the Lieb lattice and displays the properties of both.
  The super-honeycomb lattice also represents a hybrid fermionic and bosonic system, which is rarely seen in nature.
  By choosing the phases of input beams properly, the flat-band mode of the super-honeycomb will be excited
  and the input beams will exhibit strong localization during propagation.
  On the other hand, if the modes of Dirac cones of the super-honeycomb lattice are excited, one will observe conical diffraction.
  Furthermore, if the input beam is properly chosen to excite a sublattice of the super-honeycomb lattice and the modes of Dirac cones with different pseudospins, e.g., the three-beam interference pattern, the pseudospin-mediated vortices will be observed.
\end{abstract}

\maketitle


\section{Introduction}\label{introduction}

In the last decade, research in photonic crystals and optical waveguides arrays has attracted a lot of attention \cite{lederer.pr.463.1.2008,longhi.lpr.3.243.2009,kartashov.rmp.83.247.2011,garanovich.pr.518.1.2012}.
Recently, flat bands \cite{vicencio.prl.114.245503.2015,mukherjee.prl.114.245504.2015} and Dirac cones \cite{rechtsman.nature.496.196.2013,zhang.lpr.9.331.2015} were observed in the band structure of certain optical lattices
and promptly acquired a special notice.
A flat band means that the bandwidth of the band is zero, so that eigenmodes of the flat band are highly degenerate,
which can be explored to study strong correlation problems \cite{sun.prl.106.236803.2011}.
Light that excites a flat-band mode will be strongly localized during propagation,
because both the first-order and the second-order derivatives of a flat band are zero.
Due to such strong localization and nondiffracting properties, flat-band materials, e.g., Lieb lattice and kagome lattice photonic crystals,
are broadly utilized for distortion-free image transmission,
lossless optical information, and light localization \cite{xia.ol.41.1435.2016,zong.oe.24.8877.2016}.
It should be noted that there exist many flat-band materials and models \cite{deng.jssc.176.412.2003,wu.prl.99.070401.2007,tasaki.epjb.64.365.2008,crespi.njp.15.013012.2013,rechtsman.np.7.153.2013,jacqmin.prl.112.116402.2014,zhang.lpr.9.331.2015,mukherjee.ol.41.5443.2015,baboux.prl.116.066402.2016,khomeriki.prl.116.245301.2016,weimann.ol.41.2414.2016},
but the models based on the Lieb lattice \cite{apaja.pra.82.041402.2010,goldman.pra.83.063601.2011,nita.prb.87.125428.2013,guzman-silva.njp.16.063061.2014,taie.sa.1.1500854.2015,zhang.rrp.68.230.2016} are probably  the simplest.
Enlightened by the properties of a simple Lieb lattice, novel face-centered square lattices (i.e., Lieb-I and Lieb-II lattices) that possess more than one flat band were introduced recently \cite{zhang.arxiv1605.04389.2016}.
It is worth mentioning that these lattices are free of the ``pseudomagnetic effect'', as discussed in the previous literature \cite{vidal.prl.85.3906.2000,rechtsman.np.7.153.2013,longhi.ol.39.5892.2014,mukherjee.arxiv1604.05612}.

Concerning Dirac cones, which are amply discussed in the graphene or the honeycomb lattice (i.e., the photonic graphene) \cite{neto.rmp.81.109.2009,singha.science.332.1176.2011,tarruell.nature.483.302.2012,gomes.nature.483.306.2012,polini.nn.8.625.2013,jotzu.nature.515.237.2014,jacqmin.prl.112.116402.2014},
the dispersion relation in the vicinity of a Dirac cone is linear.
Hence, the first-order derivative close to the Dirac point is constant and the second-order derivative is zero,
which indicates that light exciting the modes of the Dirac cone will undergo a conical diffraction during propagation \cite{peleg.prl.98.103901.2007,ablowitz.pra.79.053830.2009,ablowitz.pra.82.013840.2010,ablowitz.siam.73.1959.2013}.
The intensity distribution then forms a circular ring, the radius of which is increasing with propagation, while the width remains the same.
To excite a mode of the Dirac cone, two methods are usually adopted to prepare the input beam \cite{ablowitz.pra.79.053830.2009}:
(1) Numerically obtain the Bloch modes of the Dirac cone and multiply the modes with a wide Gaussian beam;
(2) Find the location of the Dirac cone in the first Brillouin zone, produce the plane waves accordingly, and then multiply these waves with a wide Gaussian beam.

Another interesting phenomenon connected with the honeycomb lattice is the pseudospin, which was
theoretically predicted \cite{mecklenburg.prl.106.116803.2011,trushin.prl.107.156801.2011} and experimentally observed recently \cite{song.nc.6.6272.2015,song.2dm.2.034007.2015}.
It is important to note that the pseudospin arises from the sublattice in space and can be totally transformed into the angular momentum.
Conical diffraction and pseudospin were also investigated in the Lieb lattice \cite{leykam.pra.86.031805.2012,diebel.prl.116.183902.2016}.

Here, we investigate whether there is a kind of lattice that possess the band structure of both the honeycomb lattice and the Lieb lattice.
Considering the structure of a Lieb lattice, we introduce a novel honeycomb lattice by inserting another site in-between two nearest-neighbor (NN) sites of a honeycomb lattice.
We call such a lattice the \textit{super-honeycomb lattice}, which was also preliminarily studied more than twenty years ago \cite{shima.prl.71.4389.1993,aoki.prb.54.R17296.1996}.
The super-honeycomb lattice has 5 sites in the unit cell, and there will be 5 bands according to the tight-binding approximation method.
Our investigation will demonstrate that the band structure of a super-honeycomb lattice possesses a flat band and Dirac cones simultaneously,
so that one can observe both strong localization and conical diffraction in such a lattice.
By alternatively exciting the sublattices, a pseudospin-mediated vortex can be observed in the super-honeycomb lattice.
To the best of our knowledge, such transport properties in a super-honeycomb lattice were not reported before,
and will be studied in this paper.
We would like to note that optical lattices of this kind can be conveniently obtained by the so-called femtosecond laser writing technique \cite{davis.ol.21.1729.1996,szameit.jpb.43.163001.2010,plotnik.nm.13.57.2014,maczewsky.arxiv1605.03877}.

It should be mentioned that this is not the first investigation of the super-honeycomb lattice.
In Ref. \cite{lan.prb.85.155451.2012}, the authors showed that
the super-honeycomb lattice (which could also be denoted as the edge-centered honeycomb lattice) can support simultaneously spin-1/2 and spin-1 Dirac-Weyl fermions.
The connections among the super-honeycomb lattice, the honeycomb lattice and the kagome lattice were also revealed there.
However, the methods of investigation and the aspects of problems investigated are different here and there.
Thus, we have adopted a different research avenue from \cite{lan.prb.85.155451.2012},
to investigate transport properties of the photonic super-honeycomb lattice.
Our research has generated novel results that were not reported in \cite{lan.prb.85.155451.2012}.

The paper is organized as follows.
In Sec. \ref{theory} we introduce the mathematical model and display the schematic of the super-honeycomb lattice.
By applying the tight-binding method (only considering the NN hopping), the band structure of the super-honeycomb lattice is obtained.
Transport properties are discussed in Secs. \ref{flat} and \ref{transport}.
In particular, Sec. \ref{flat} discusses light localization due to a flat band.
We show that both the intensity and the phase of light remain the same during propagation, if the flat-band mode is excited,
or, if not, they undergo discrete diffraction.
In Sec. \ref{cone}, the conical diffraction of light that excites the Dirac cone is presented.
We discuss the generation of pseudospin-mediated vortices by exciting a sublattice of the super-honeycomb lattice in Sec. \ref{pesudospin}.
We conclude the paper in Sec. \ref{conclusion}.

\section{Mathematical model} \label{theory}

\begin{figure}[htpb]
\centering
  \includegraphics[width=0.5\columnwidth]{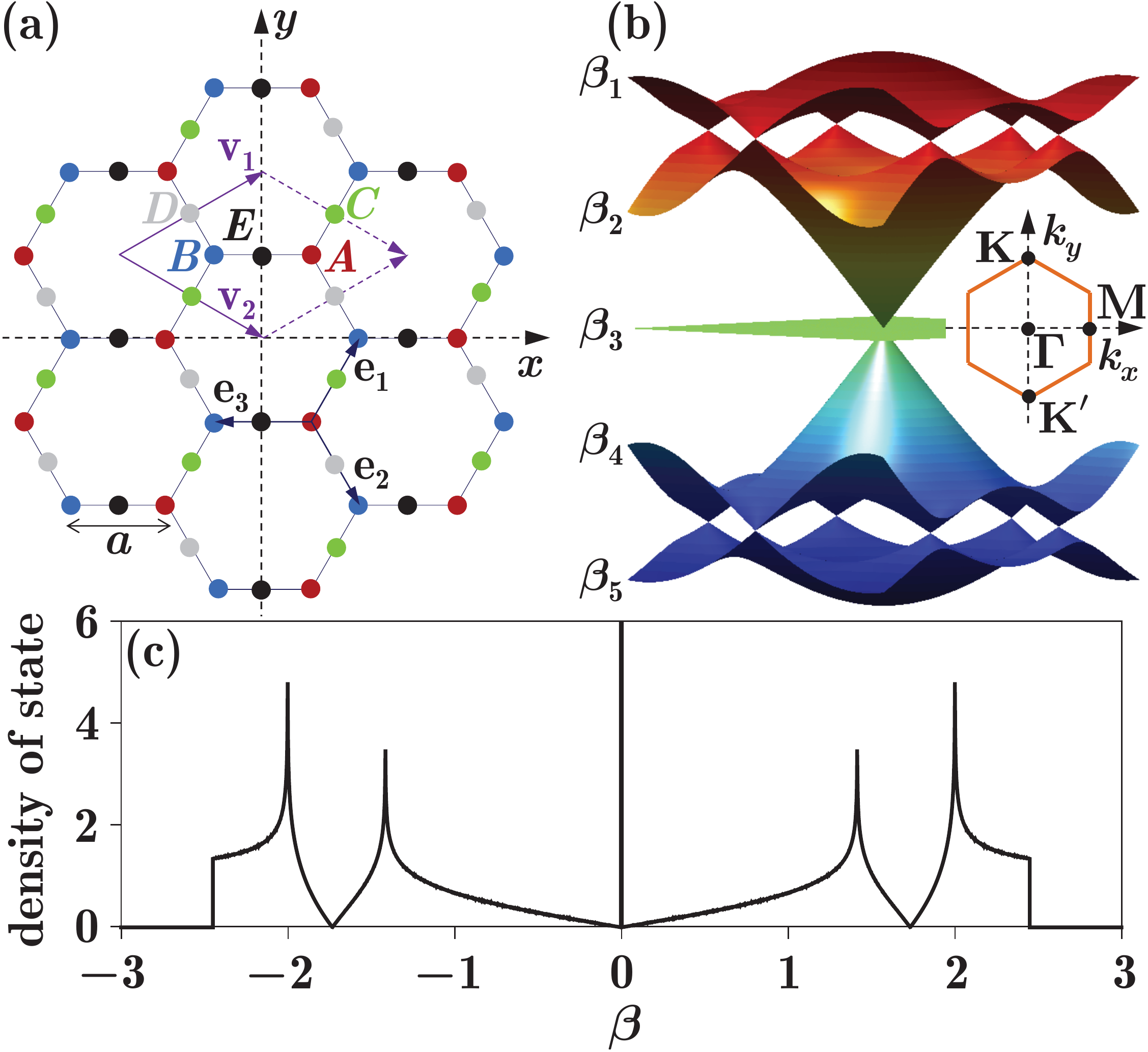}
  \caption{(a) Schematic of the super-honeycomb lattice with 5 sites marked as \textit{A}, \textit{B}, \textit{C}, \textit{D} and \textit{E} in the unit cell.
  $\mathbf{v_1}$ and $ \mathbf{v_2}$ are the basis vectors in real space.
  (b) Band structure; from top to bottom, the bands are $\beta_{1\sim5}$, respectively.
  Inset shows the first Brillouin zone and the high symmetry points $\bf \Gamma$, \textbf{M}, \textbf{K} and ${\bf K}'$.
  (c) The corresponding density of states.}
  \label{fig1}
\end{figure}

The geometry of the super-honeycomb lattice is shown in Fig. \ref{fig1}(a).
Different from the honeycomb lattice, there are 5 sites in the unit cell of a super-honeycomb lattice.
Similar to the honeycomb lattice, there are two sublattices, formed by three $A$ sites and three $B$ sites, respectively.
Even though $C$, $D$ and $E$ sites can form another two triangles,
the properties of the three sites are different (they are not equivalent to each other).
Therefore, there are 5 sublattices in a super-honeycomb lattice.
By adopting the tight-binding method and assuming that hopping only occurs between the NN sites,
the propagation of light in this discrete model obeys the discrete coupled Schr\"odinger equations \cite{szameit.pra.84.021806.2011,zhang.arxiv1605.04389.2016}:
\begin{subequations}\label{eq1}
\begin{equation}\label{eq1a}
i\frac{{\partial {A_m}}}{{\partial z}} = t\sum\limits_{{{\bf{R}}_m},{{\bf{e}}_i}} {({C_m} + {D_m} + {E_m})},
\end{equation}
\begin{equation}\label{eq1b}
i\frac{{\partial {B_m}}}{{\partial z}} = t\sum\limits_{{{\bf{R}}_m},{{\bf{e}}_i}} {({C_m} + {D_m} + {E_m})},
\end{equation}
\begin{equation}\label{eq1c}
i\frac{{\partial {C_m}}}{{\partial z}} = t\sum\limits_{{{\bf{R}}_m},{{\bf{e}}_i}} {({A_m} + {B_m})},
\end{equation}
\begin{equation}\label{eq1d}
i\frac{{\partial {D_m}}}{{\partial z}} = t\sum\limits_{{{\bf{R}}_m},{{\bf{e}}_i}} {({A_m} + {B_m})},
\end{equation}
\begin{equation}\label{eq1e}
i\frac{{\partial {E_m}}}{{\partial z}} = t\sum\limits_{{{\bf{R}}_m},{{\bf{e}}_i}} {({A_m} + {B_m})},
\end{equation}
\end{subequations}
where $\textit{z}$ is the longitudinal distance and $\textbf{R}_m$ is the position of the $m$th unit cell.
Thus, it is assumed that a photonic crystal consisting of an array of optical waveguides is arranged in the form of super-honeycomb lattice and that
light propagates perpendicular to the lattice, along the waveguides.
For simplicity, we set the distance between sites $A$ and $B$ to be $a=1$, and the hopping strength to be $t=1$.
${{\bf{e}}_{1,2,3}}$ are the 3 vectors representing the hopping directions,
with components ${{\bf{e}}_1} = (1/2,\sqrt 3 /2)$, ${{\bf{e}}_2} = (1/2, - \sqrt 3 /2)$, and ${{\bf{e}}_3} = ( - 1,0)$;
${\bf v}_{1,2}$ are the Bravais primitive vectors, with components ${\bf v}_1=(3/2,\sqrt{3}/2)$ and ${\bf v}_2=(3/2,-\sqrt{3}/2)$.

In order to find solutions of Eqs. (\ref{eq1}), we introduce an ansatz of the form \cite{szameit.pra.84.021806.2011}
\begin{equation}\label{eq2}
\begin{split}
{A_m} = & ~{A_{\bf{k}}}\exp [i(\beta z + {{\bf{R}}_m} \cdot {\bf{k}})],\\
{B_m} = & ~{B_{\bf{k}}}\exp [i(\beta z + {{\bf{R}}_m} \cdot {\bf{k}})],\\
{C_m} = & ~{C_{\bf{k}}}\exp [i(\beta z + {{\bf{R}}_m} \cdot {\bf{k}})],\\
{D_m} = & ~{D_{\bf{k}}}\exp [i(\beta z + {{\bf{R}}_m} \cdot {\bf{k}})],\\
{E_m} = & ~{E_{\bf{k}}}\exp [i(\beta z + {{\bf{R}}_m} \cdot {\bf{k}})].
\end{split}
\end{equation}
Plugging Eq. (\ref{eq2}) into Eq. (\ref{eq1}), after some algebra, one obtains:
\begin{equation}\label{eq3}
H_{TB} |\beta ,\bf{k}\rangle  = \beta |\beta ,\bf{k} \rangle,
\end{equation}
in which
\begin{equation}\label{eq4}
|\beta,{\bf k}\rangle=
\begin{bmatrix}
A_{\bf{k}}\\
B_{\bf{k}}\\
C_{\bf{k}}\\
D_{\bf{k}}\\
E_{\bf{k}}
\end{bmatrix},
\end{equation}
and the tight-binding Hamiltonian of the system is:
\begin{equation}\label{eq5}
H_{TB} = -
\begin{bmatrix}
  0          & 0               & H_{13}             & H_{14}           & H_{15}             \\
  0          & 0               & H_{13}^*           & H_{14}^*         & H_{15}^*             \\
  H_{13}^*   & H_{13}          & 0                  & 0                & 0 \\
  H_{14}^*   & H_{14}          & 0                  & 0                & 0  \\
  H_{15}^*   & H_{15}          & 0                  & 0                & 0
\end{bmatrix},
\end{equation}
where ${H_{13}}=\exp [  i({k_x}/4 + \sqrt 3 {k_y}/4)]$,
${H_{14}}=\exp [  i({k_x}/4 - \sqrt 3 {k_y}/4)]$,
${H_{15}}=\exp ( - i{k_x}/2)$; here $(\bullet)^*$ denotes the complex conjugate of $(\bullet)$.
Obviously, Eq. (\ref{eq5}) is a Hermitian matrix,
so that one can solve for five real eigenvalues $\beta_{1\sim 5}$ that represent the dispersive relation, i.e., the band structure.
It is not difficult to demonstrate that $\beta_1(k_x,k_y)=-\beta_5(k_x,k_y)$, $\beta_2(k_x,k_y)=-\beta_4(k_x,k_y)$, and $\beta_{3}=0$.
Since $\beta_3$ is independent of $k_x$ or $k_y$, it corresponds to a complete flat band in the first Brillouin zone.
The corresponding eigenstate of the flat band is
\begin{align}\label{eq6}
|\beta_3, \bf{k}\rangle  =
\begin{bmatrix}
0\\
0\\
{\sin \left( {3{k_x}/4 - \sqrt 3 {k_y}/4} \right)}\\
{ - \sin \left( {3{k_x}/4 + \sqrt 3 {k_y}/4} \right)}\\
{\sin \left( {\sqrt 3 {k_y}/2} \right)}
\end{bmatrix},
\end{align}
which means that the flat-band mode has vanishing amplitude on sublattices \textit{A} and \textit{B}.

Figure \ref{fig1}(b) shows the numerical dispersion relation of the super-honeycomb lattice in the first Brillouin zone.
One finds that the bands are symmetric about the flat band,
and that there are 6 Dirac cones between $\beta_1$ and $\beta_2$, as well as between $\beta_4$ and $\beta_5$ at the corners of the first Brillouin zone,
which is similar to the Dirac cones in the honeycomb lattice.
Intersected by the flat band, there is another Dirac cone, which is similar to that in the Lieb lattice, but located at the origin of the first Brillouin zone.
We note that typically there are two symmetric sites in the first Brillouin zone that are called $\bf K$ and ${\bf K}'$.
As shown in the inset in Fig. \ref{fig1}(b), the locations of $\bf K$ and ${\bf K}'$ are $(0,\pm4\pi/3\sqrt{3})$.
Due to the symmetry of the band structure, there will be a state $|-\beta_1, \mathbf{k}\rangle$ for each $|\beta_1, \mathbf{k}\rangle$,
and $|-\beta_2, \mathbf{k}\rangle$ for each $|\beta_2, \mathbf{k}\rangle$, therefore the particle-hole symmetry exists in the super-honeycomb lattice.
In fact, the criteria for the existence of such a symmetry is that there is an operator $\hat O$ which anticommutes with the Hamiltonian \cite{leykam.pra.86.031805.2012}.
It is easy to verify that the operator $\hat O$ for this case is $\hat O = {\rm diag}(1,1,-1,-1,-1)$.
Looking at the band structure shown in Fig. \ref{fig1}(b), one can draw a conclusion that the super-honeycomb lattice is a hybrid of the honeycomb lattice and the Lieb lattice,
because there are simultaneously ``pseudospin-1/2'' Dirac cones
and ``pseudospin-1'' Dirac cones in it, which are present in the honeycomb lattice \cite{mecklenburg.prl.106.116803.2011,trushin.prl.107.156801.2011,song.nc.6.6272.2015,song.2dm.2.034007.2015}
and in the Lieb lattice \cite{leykam.pra.86.031805.2012,diebel.prl.116.183902.2016,leykam.aipx.1.101.2016} separately.

We also display the corresponding density of states \cite{neto.rmp.81.109.2009} of the band structure
in Fig. \ref{fig1}(c), to classify and distinguish different pseudospins
embedded in the super-honeycomb lattice.
One can clearly see a $\delta$-function singularity at $\beta=0$ (due to the flat band),
therefore the corresponding Dirac cone is bosonic and $s=1$ (where $s$ represents the pseudospin) \cite{leykam.aipx.1.101.2016}.
Whereas for the other two Dirac cones the density of states is zero, so they are fermionic and $s=1/2$.

Now, we present a brief analysis of the pseudospins connected with the Dirac cones at $\bf K$ and $\bf \Gamma$.
Around the Dirac points (\textbf{K} or $\bf \Gamma$) in the band structure, the Hamiltonian in Eq. (\ref{eq5}) can be rewritten as
\begin{equation}\label{eq7}
H_{\bf p} = -
\begin{bmatrix}
  0          & 0               & \gamma H_{13,\textbf{p}}             & \gamma^* H_{14,\textbf{p}}           & H_{15,\textbf{p}}             \\
  0          & 0               & \gamma^* H_{13,\textbf{p}}^*           & \gamma H_{14,\textbf{p}}^*         & H_{15,\textbf{p}}^*             \\
  \gamma^* H_{13,\textbf{p}}^*   & \gamma H_{13,\textbf{p}}          & 0                  & 0                & 0 \\
  \gamma H_{14,\textbf{p}}^*   & \gamma^* H_{14,\textbf{p}}          & 0                  & 0                & 0  \\
  H_{15,\textbf{p}}^*   & H_{15,\textbf{p}}          & 0                  & 0                & 0
\end{bmatrix},
\end{equation}
where ${\bf p}=[p_x,p_y]$ is the displacement from the Dirac point,
$H_{13,\textbf{p}}=1+i({p_x}/4 + \sqrt 3 {p_y}/4)$,
$H_{14,\textbf{p}}=1+i({p_x}/4 - \sqrt 3 {p_y}/4)$,
$H_{15,\textbf{p}}=1 - i{p_x}/2$,
and $\gamma$ is the coefficient that is $\exp(i\pi/3)$ for the Dirac point at \textbf{K} and 1 for the Dirac point at $\bf \Gamma$.
One can calculate that the dispersion relation is $\beta=\sqrt{3}(1\pm|{\bf p}|/4)$ around \textbf{K} and $\beta=\pm\sqrt{3}|{\bf p}|/2$ around $\bf \Gamma$.
Considering that $\nabla_{\bf p}\beta$ determines the group velocity of the beam,
the speed of beam spreading around $\bf \Gamma$ is twice that of $\bf K$; this was also found in Ref. \cite{lan.prb.85.155451.2012}.

We first consider the Hamiltonian $H_{\bf p}$ around the $\bf \Gamma$ point;
one finds that $[H_{\bf p},{\hat L}_z]\neq0$.
So, there must be some additional angular momentum missing, which is the so-called pseudospin.
If we define the pseudospin operator as ${\hat S}_z$, the total angular momentum can be written as ${\hat J}_z = {\hat L}_z + {\hat S}_z$,
which meets the condition $[H_{\bf p},{\hat J}_z]=0$.
Indeed, one finds that
\begin{align}\label{eq8}
2\sqrt{3} S_x =
\begin{bmatrix}
 0 & 0 & i & i & -2i \\
 0 & 0 & -i & -i & 2i \\
 -i & i & 0 & 0 & 0 \\
 -i & i & 0 & 0 & 0 \\
 2i & -2i & 0 & 0 & 0
\end{bmatrix},
\end{align}
\begin{align}\label{eq9}
2 S_y =
\begin{bmatrix}
 0 & 0 & i & -i & 0 \\
 0 & 0 & -i & i & 0 \\
 -i & i & 0 & 0 & 0 \\
 i & -i & 0 & 0 & 0 \\
 0 & 0 & 0 & 0 & 0
\end{bmatrix},
\end{align}
\begin{align}\label{eq10}
\sqrt{3} S_z = &
\begin{bmatrix}
0 & 0 & 0 & 0 & 0\\
0 & 0 & 0 & 0 & 0\\
0 & 0 & 0 & i & -i\\
0 & 0 & -i & 0 & i\\
0 & 0 & i & -i & 0\\
\end{bmatrix},
\end{align}
which satisfies the angular momentum algebra conditions $[S_j,S_k]=i\epsilon_{jkl}S_l$ and $[S^2,S_k]=0$ with $S^2=\sum_k S_k^2$.
Diagonalizing $S_z$, one obtains
\begin{subequations}\label{eq11}
\begin{align}
\label{eq11a}|S_z=0\rangle =  [1,~0,~0,~0,~0]^T,
\end{align}
\begin{align}
\label{eq11b}|S_z=0\rangle =  [0,~1,~0,~0,~0]^T,
\end{align}
\begin{align}
\label{eq11c}|S_z=0\rangle =  [0,~0,~1,~1,~1]^T,
\end{align}
\begin{align}
\label{eq11d}|S_z=1\rangle =  [0,~0,~\exp(-i2\pi/3),~\exp(i2\pi/3),~1]^T,
\end{align}
\begin{align}
\label{eq11e}|S_z=-1\rangle =  [0,~0,~\exp(i2\pi/3),~\exp(-i2\pi/3),~1]^T.
\end{align}
\end{subequations}
Hence, for the $|S_z=0\rangle$ state, there are two cases: vanishing amplitude on \textit{A} and \textit{B},
and non-vanishing amplitude only on \textit{A} or \textit{B}.
Whereas for the $|S_z=\pm1\rangle$ states, the amplitudes on \textit{A} and \textit{B} vanish,
and the amplitudes on \textit{C}, \textit{D} and \textit{E} have a relative phase difference of $2\pi/3$.
We determine that the pseudospin corresponding to the Dirac cone at $\bf \Gamma$ can be 0 or 1.

There is also pseudospin at the Dirac point $\bf K$ in the super-honeycomb lattice;
however, the pseudospin matrix corresponding to this case is nontrivial
\footnote{According to $[H_\textbf{p},\hat{S}_z]=-[H_\textbf{p},\hat{L}_z]$, one can obtain 25 linear equations,
which can be rewritten in matrix format: $\mathcal{M}\mathbf{S}=\mathbf{L}$,
in which $\mathcal{M}$ is the coefficient matrix, and
$\mathbf{S}$ and $\mathbf{L}$ are $25^2\times1$ vectors constructed from matrix $S_z$ and $-[H_\textbf{p},\hat{L}_z]$, respectively.
One can easily verify that the rank of the matrix $\mathcal{M}$ is 20 and the rank of the corresponding augmented matrix is 21.
Therefore, there is no solution to the equation $\mathcal{M}\mathbf{S}=\mathbf{L}$.}.
At least, one clear point is that the appearance of the Dirac cones is only due to the sublattices \textit{A} and \textit{B}.
One can consider the 2-band Hamiltonian that describes the Dirac cone at \textbf{K},
to find that the pseudospin is 1/2 \cite{sun.prl.103.046811.2009,lan.prb.85.155451.2012,leykam.phd.2015},
which is similar to that in the honeycomb lattice or the kagome lattice;
however, the relations among \textit{C}, \textit{D} and \textit{E} are not changed.
Therefore, we just discuss the intensity distributions on sublattices \textit{A} and \textit{B}
when analyzing the pseudospin-mediated vortex generation associated with the Dirac cone at \textbf{K} in Sec. \ref{pesudospin};
those on sublattices \textit{C}, \textit{D} and \textit{E} are not involved in the indistinct projections.


\section{Strong localization due to the flat band} \label{flat}

\begin{figure}[htpb]
 \centering
 \includegraphics[width=0.5\columnwidth]{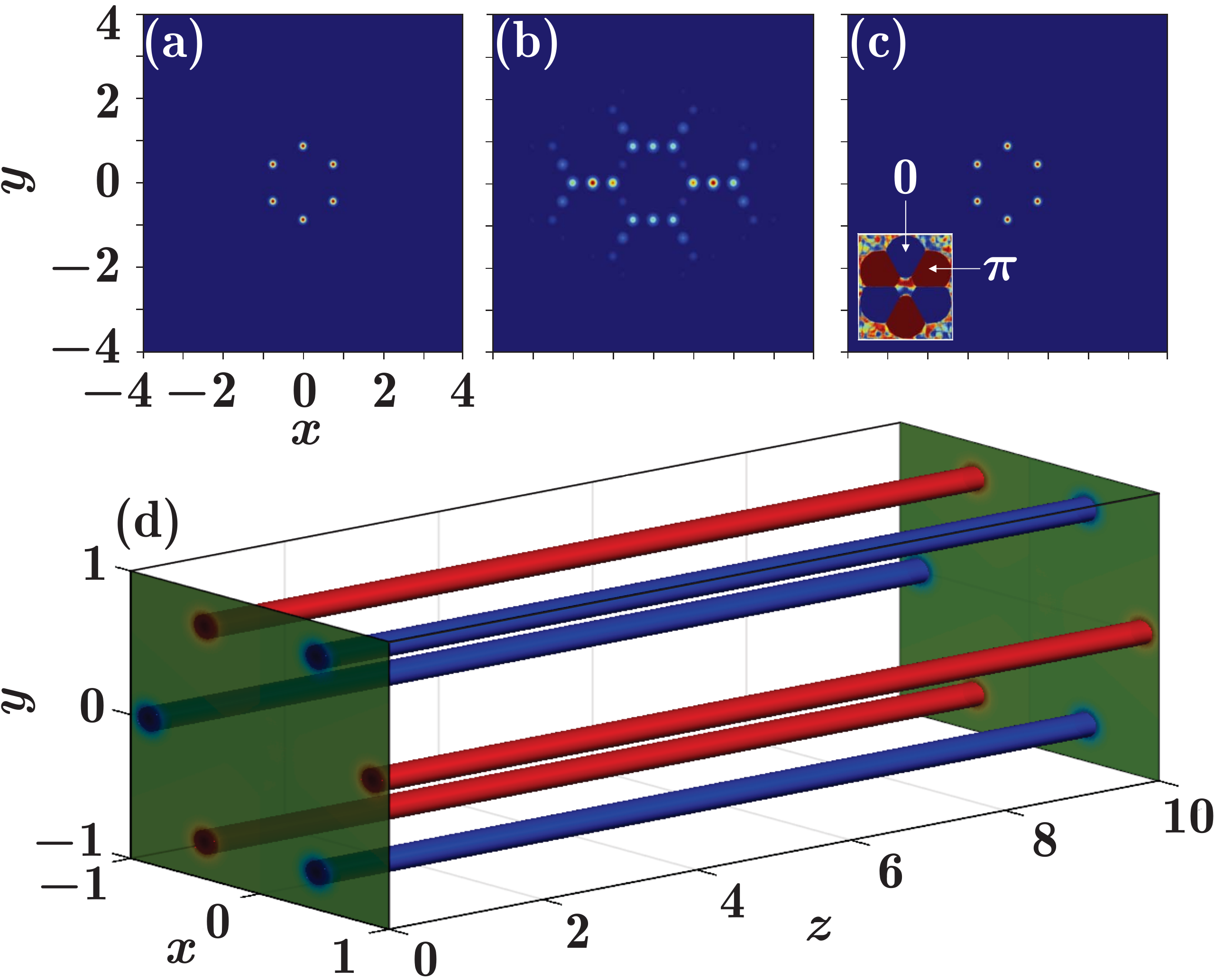}
  \caption{(a) Intensity of the input beam that contains six Gaussian beams pointed at the sublattices \textit{C}, \textit{D} and \textit{E}.
  (b) Output intensity of the input beam if the six Gaussian beams are neither in-phase nor out-of-phase. Discrete diffraction happens during propagation.
  (c) Same as (b), but for the out-of-phase case. Strong localization is observed during propagation. Inset: phase of the output beam.
  (d) Iso-surface plot of the propagation corresponding to (c).}
  \label{fig2}
  \end{figure}

It has been proved that if light excites the mode of the flat band of the Lieb lattice or the kagome lattice,
it will be strongly localized during propagation in the corresponding waveguides \cite{vicencio.jo.16.015706.2014,vicencio.prl.114.245503.2015,mukherjee.prl.114.245504.2015,xia.ol.41.1435.2016,zong.oe.24.8877.2016}.
In this section, we investigate the localization of the flat-band mode in the super-honeycomb lattice, according to Eqs. (\ref{eq1}).
Based on Eq. (\ref{eq6}), the flat-band mode has vanishing amplitude on sublattices \textit{A} and \textit{B},
so we launch the input beam --- six Gaussian beams --- into sublattices \textit{C}, \textit{D} and \textit{E}, as presented in Fig. \ref{fig2}(a).
First, we assume that the six Gaussian beams are neither in-phase (according to Eq. (\ref{eq11a}), conical diffraction will then be obtained) nor out-of-phase.
On this occasion, the input beam exhibits discrete diffraction during propagation in the lattice waveguides, as shown in Fig. \ref{fig2}(b),
which means that neither the in-phase nor the out-of-phase input beam excites the flat-band mode.

On the other hand, if the six beams are out-of-phase, the flat-band mode will be excited \cite{mukherjee.prl.114.245504.2015,xia.ol.41.1435.2016,zhang.arxiv1605.04389.2016}
and the strong localization during propagation will be observed, as shown in Fig. \ref{fig2}(c).
As expected, once the flat-band mode is excited, the beam is localized ---
the output intensity is the same as the initial in Fig. \ref{fig2}(a).
Concerning the phase of the output beam, as shown in the inset in Fig. \ref{fig2}(c),
the phase is also preserved very well, i.e., the six peaks are still out-of-phase.
The physical explanation to this phenomenon is quite natural.
Since the band width of the flat band is 0, the first-order derivative (corresponding to the group velocity of the beam)
as well as the second-order derivative (corresponding to the diffraction of the beam) of the flat band is 0.
As a result, the beam will not diffract during propagation and it will remain the same --- an indication of strong localization.

To see the strong localization more clearly, in Fig. \ref{fig2}(d) we show the whole propagation of the input beam used in Fig. \ref{fig2}(c),
by connecting the corresponding iso-surfaces (which form tubes) of the beam during propagation.
In Fig. \ref{fig2}(d), the amplitudes of the input beam and the output beam are displayed at the input face and the output face, respectively,
and there is a $\pi$-phase difference between the red tubes and the blue ones.
Again, one observes that both the amplitude and the phase of the beam do not change during propagation ---
both the width and the color of the tubes remain unchanged.
Comparing Fig. \ref{fig2}(b) with Fig. \ref{fig2}(c),
one comes to the conclusion that not only the amplitude but also the phase plays an important role in
the excitation of the flat-band mode.

\section{Conical diffraction and pseudospin due to Dirac cones}\label{transport}

\subsection{Conical diffraction} \label{cone}

After discussing the localization caused by the flat band in the super-honeycomb lattice,
we now analyze the influence of Dirac cones on the beam propagation dynamics.
As already mentioned, there are two kinds of Dirac cones in the band structure shown in Fig. \ref{fig1}(b),
corresponding to pseudospin 1/2 (at \textbf{K} and ${\bf K}'$) and pseudospin 1 (at $\bf\Gamma$), respectively.
The group speed of the beam due to the former Dirac cones is half of that due to the latter.
We first discuss the first kind of Dirac cones and later the latter kind.

\begin{figure}[htpb]
\centering
 \includegraphics[width=0.5\columnwidth]{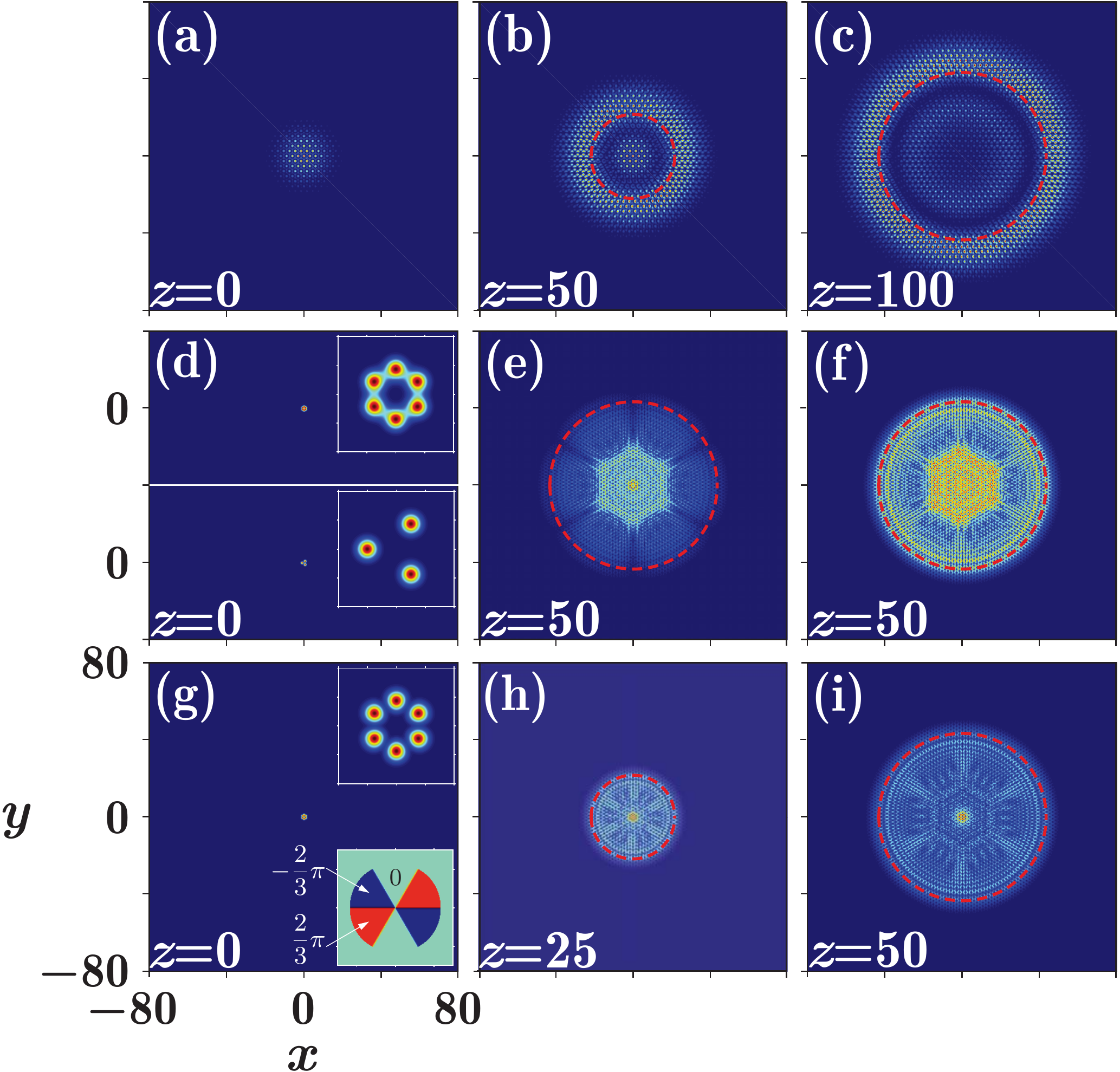}
  \caption{Conical diffraction during propagation.
  (a)-(c) Intensity distributions of the beam at certain propagation distances, displayed in the left-bottom corner of each panel.
  Six wide Gaussian beams multiplied with the plane waves at the six Dirac points are launched into the sublattices \textit{A} and \textit{B}.
  (d) Square root of the amplitude of the pseudospin-0 input.
  Upper panel: six in-phase narrow Gaussian beams are launched into the sublattices \textit{C}, \textit{D} and \textit{E}.
  Bottom panel: three in-phase narrow Gaussian beams are launched into the sublattice \textit{A}.
  Insets show the magnified beams.
  (e) and (f) Square root of the amplitude of the output beam corresponding to the upper and lower inputs in (d), respectively.
  (g)-(i) Square root of the amplitudes of the pseudospin-1 beam.
  Six narrow Gaussian beams with a $2\pi/3$ phase difference between adjacent beams
  (upper inset: magnified beam, bottom panel: phase distribution),
  are launched into the sublattices \textit{C}, \textit{D} and \textit{E}.
  Dashed circles represent the analytical radius of the conical diffraction.}
  \label{fig3}
  \end{figure}

In the neighbourhood of a Dirac point, the shape of the band structure is cone-like.
So, the corresponding first-order derivative is constant, while the second-order derivative is zero.
That is, the beam will linearly spread during propagation, but the beam width will not change.
By definition, this is the conical diffraction \cite{peleg.prl.98.103901.2007,ablowitz.pra.79.053830.2009,ablowitz.pra.82.013840.2010,leykam.pra.86.031805.2012,song.nc.6.6272.2015,song.2dm.2.034007.2015,diebel.prl.116.183902.2016,leykam.aipx.1.101.2016,zhang.sr.6.23645.2016,zhang.lpr.10.526.2016}.
To see conical diffraction during propagation in the super-honeycomb lattice,
we assume that the input beam is constructed according to the second method mentioned in the introduction, as shown in Fig. \ref{fig3}(a).
Figures \ref{fig3}(b) and \ref{fig3}(c) depict the intensity distribution of the beam at certain distances,
at which the beam indeed shows conical diffraction, as expected.
The dashed circles in the panels correspond to the analytical locations of the conical rings, obtained according to the group speed of the beam.
One finds that the analytical predictions and numerical results agree with each other very well.
Similar to previous research, there are two bright rings separated by a dark ring --- the Poggendorff's dark ring \cite{berry.po.50.13.2007}
--- and the intensity of the outer ring is higher than the intensity of the inner one.

To excite the mode of the Dirac cone at $\bf \Gamma$, the input beam can be prepared in two different ways, according to Eq. (\ref{eq11}):
one is the pseudospin-0 case, the other the pseudospin-1 case.
For the pseudospin-0 case, we first launch six narrow Gaussian beams into the sublattices \textit{C}, \textit{D} and \textit{E} according to Eq. (\ref{eq11c});
the square root of the beam amplitude is shown in the upper panel in Fig. \ref{fig3}(d),
and the corresponding output is shown in Fig. \ref{fig3}(e).
We use the square root of the beam amplitude instead of the beam intensity,
because the narrow Gaussian beams might excite the modes of other dispersive bands except the modes of Dirac cones.
This would lead to high intensity at the beam center during propagation, as seen in Fig. \ref{fig3}(e),
and make the conical diffraction hard to recognize \cite{ablowitz.pra.79.053830.2009,leykam.pra.86.031805.2012}.

If one launches three narrow Gaussian beams into the sublattice \textit{A} according to Eq. (\ref{eq11a}),
as presented in the bottom panel in Fig. \ref{fig3}(d),
the corresponding output is displayed in Fig. \ref{fig3}(f).
From Figs. \ref{fig3}(e) and \ref{fig3}(f) one can see that
the output beam indeed exhibits conical diffraction and the numerical results agree well with the analytical predictions (dashed circles).
Since the group speed of the mode of the Dirac cone at $\bf \Gamma$ is twice that of the cone at $\bf K$,
the beam radii in Figs. \ref{fig3}(e) and \ref{fig3}(f),
which are recorded at $z=50$, are the same as those in Fig. \ref{fig3}(c), which are recorded at $z=100$.
Concerning the pseudospin-1 case, we prepare the input according to Eq. (\ref{eq11d}), as shown in Fig. \ref{fig3}(g).
The inset displays the corresponding phase distribution of the input beam. If the input beam is prepared according to Eq. (\ref{eq11e}),
the corresponding phase distribution is opposite of that shown in the inset.
Figures \ref{fig3}(h) and \ref{fig3}(i) display the square root of the beam amplitude at certain propagation distances,
from which one observes the pseudospin-1 conical diffraction.

\begin{figure}[htpb]
\centering
 \includegraphics[width=0.5\columnwidth]{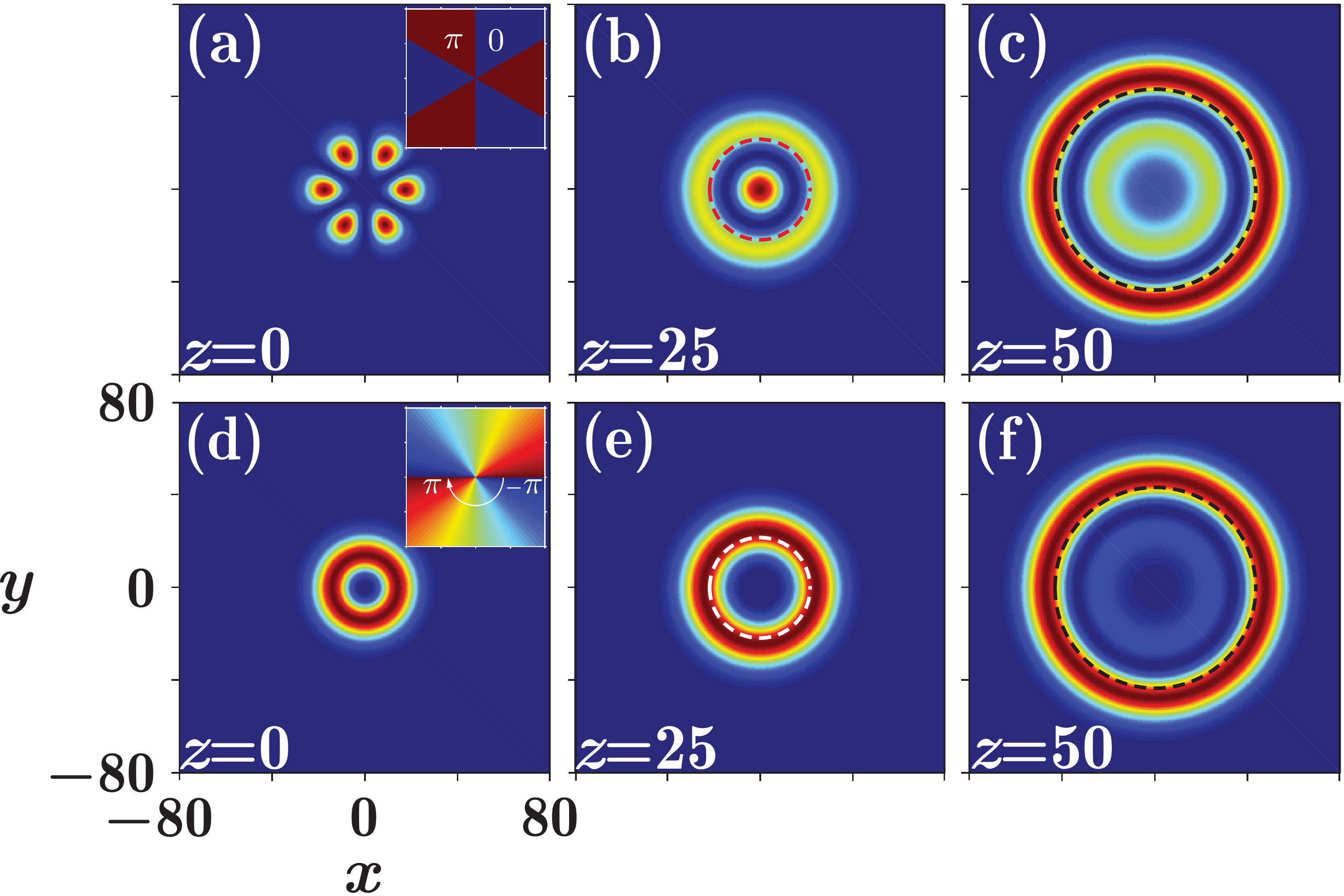}
  \caption{Conical diffraction from wide Gaussian beams.
  (a) Six wide Gaussian beams with a $\pi$-phase difference are launched into sublattices \textit{A} and \textit{B}.
  (b) and (c) Beam intensities at certain distances.
  (d) Same as Fig. \ref{fig3}(g), but for wide Gaussian beams.
  (e) and (f) Setup is as in (b) and (c).
  Insets are the phase distributions.}
  \label{fig4}
\end{figure}

We would like to point that we used narrow Gaussian beams to avoid mixing of sublattices;
if one launches wide Gaussians into a sublattice,
the other sublattices will be lighted simultaneously.
Therefore, the input condition will be destroyed,
and one typical example is the case based on Eq. (\ref{eq11c}).
However, if one simultaneously launches six wide Gaussian beams into sublattices \textit{A} and \textit{B} according to Eqs. (\ref{eq11a}) and (\ref{eq11b}),
and assumes a $\pi$-phase difference between the beams on sublattices \textit{A} and \textit{B},
the input light leakage from sublattices \textit{A} and \textit{B} onto sublattices \textit{C}, \textit{D} and \textit{E} will be suppressed, as shown in Fig. \ref{fig4}(a).
To see the conical diffraction, we just display the intensity distributions on sublattice \textit{A}, as shown in Figs. \ref{fig4}(b) and \ref{fig4}(c).
Since the input Gaussian beams are quite wide, the mode of the Dirac cone is well excited, and the conical diffraction apparently appears.
Similar method can be applied to the pseudospin-1 case, as shown in Figs. \ref{fig4}(d)-\ref{fig4}(f).

Until now, we have discussed conical diffraction due to Dirac cones with different psedudospins in the super-honeycomb lattice.
In the following subsection, we investigate the pseudospin-mediated vortices.

\subsection{Pseudospin-mediated vortices} \label{pesudospin}

It has been demonstrated before that there is pseudospin in a honeycomb lattice \cite{mecklenburg.prl.106.116803.2011,trushin.prl.107.156801.2011,song.nc.6.6272.2015,song.2dm.2.034007.2015}.
It has also been demonstrated that pseudospin also exists in the Lieb lattice \cite{leykam.pra.86.031805.2012,diebel.prl.116.183902.2016}.
It should be stressed that pseudospins in the honeycomb and Lieb lattices are different --- the former is fermionic and the latter is bosonic.
This is determined by the order of the Dirac cone in the corresponding band structure \cite{lan.prb.84.165115.2011,leykam.aipx.1.101.2016}.
As shown in Fig. \ref{fig1}(b), the Dirac cones in the band structure of the super-honeycomb lattice have different orders,
therefore both fermionic and bosonic pseudospins can be observed in the super-honeycomb lattice \cite{lan.prb.84.165115.2011}.
Indeed, as investigated above in this paper, the super-honeycomb lattice is a combination of the honeycomb lattice and the Lieb lattice.
We believe that the reason why such a lattice expresses pseudospins is that it is not a Bravais lattice,
and the total wavefunction is a direct sum of the wavefunctions of the sublattices.

\begin{figure}[htpb]
 \centering
 \includegraphics[width=0.5\columnwidth]{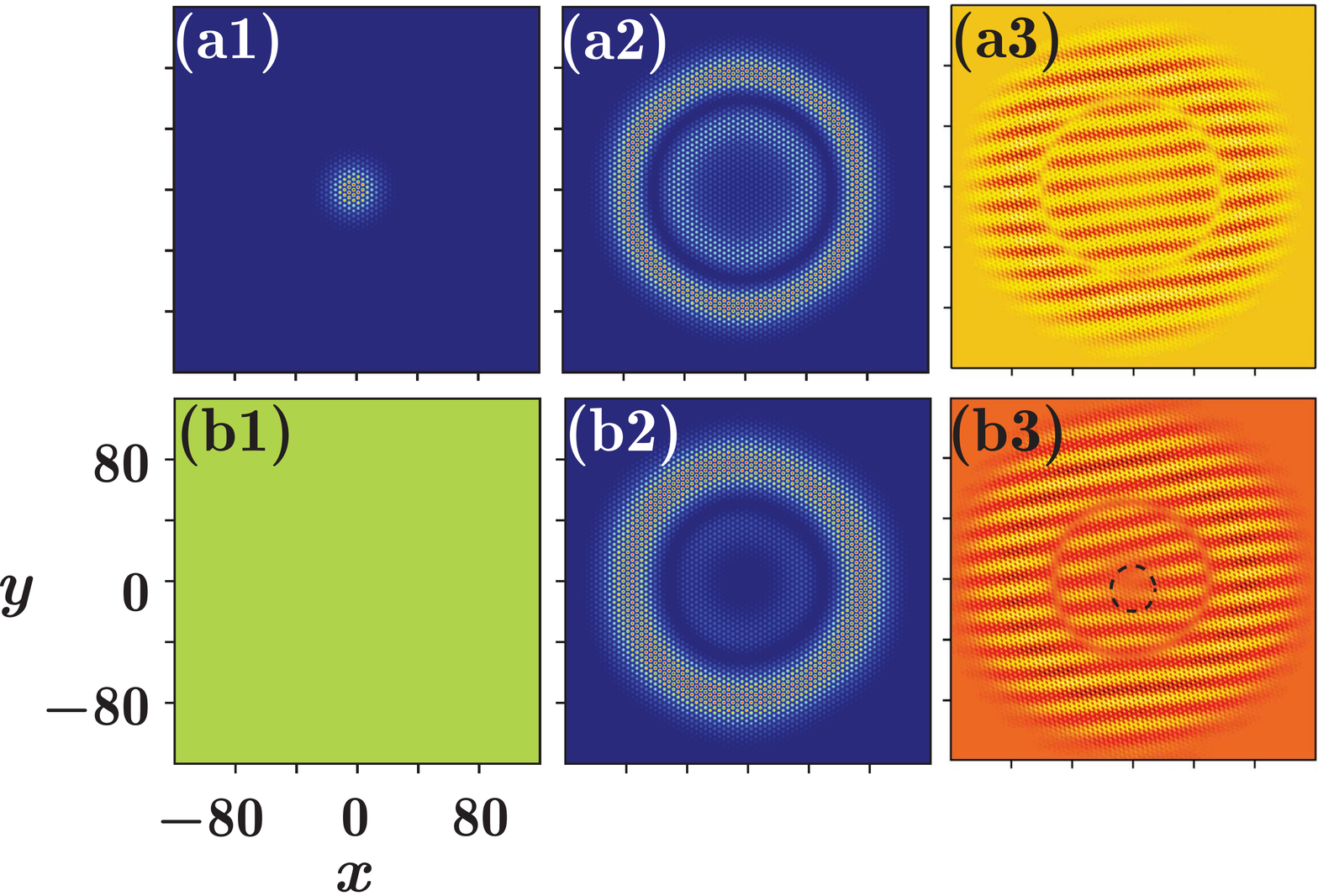}
  \caption{Intensity and phase of the optical field from numerical simulation of Eq. (\ref{eq1}),
  when only sublattice $A$ is excited using the three-beam interference pattern.
  Left column: input intensity; Middle column: output intensity; Right column: interferograms of the output beams with a tilted plane wave.
  The dashed circle shows the location of the phase singularity.}
  \label{fig5}
\end{figure}

We first consider the pseudospin-1/2 Dirac cones at \textbf{K} and ${\bf K}'$ in the super-honeycomb lattice.
To this end, we selectively launch the input beam into the sublattice \textit{A} (or \textit{B})
by using the interference pattern among three broad Gaussian beams \cite{song.nc.6.6272.2015,song.2dm.2.034007.2015}, as shown in Fig. \ref{fig5}(a1).
In Fig. \ref{fig5}(b1), the beam intensity on sublattice \textit{B} is also displayed,
which is zero --- no light beam is launched into the sublattice.
This input beam will excite the modes of the Dirac cones at $\bf K$ and ${\bf K}'$, and then exhibit conical diffraction during propagation.
As demonstrated in previous research \cite{mecklenburg.prl.106.116803.2011,trushin.prl.107.156801.2011,song.nc.6.6272.2015,song.2dm.2.034007.2015},
the pseudospin could be completely transferred into the angular momentum,
so in order to check such a transfer in the super-honeycomb lattice,
we show the output intensities on the sublattices \textit{A} and \textit{B} separately in Figs. \ref{fig5}(a2) and \ref{fig5}(b2),
and the corresponding interferograms in Figs. \ref{fig5}(a3) and \ref{fig5}(b3).
Since the sublattice \textit{A} is excited, there is no vortex generated on this sublattice, as shown in Fig. \ref{fig5}(a3).
However, as shown in Fig. \ref{fig5}(b3), a vortex is generated on sublattice \textit{B}  with a topological charge +1,
because of the bifurcation in the interferogram.
That is, when sublattice \textit{A} is excited, the psedudospin is converted to a vortex angular momentum,
and such an angular momentum will be indicated in sublattice \textit{B}.
This is same as in the previous research on a honeycomb lattice \cite{song.nc.6.6272.2015,song.2dm.2.034007.2015}.

\begin{figure}[htpb]
\centering
 \includegraphics[width=0.5\columnwidth]{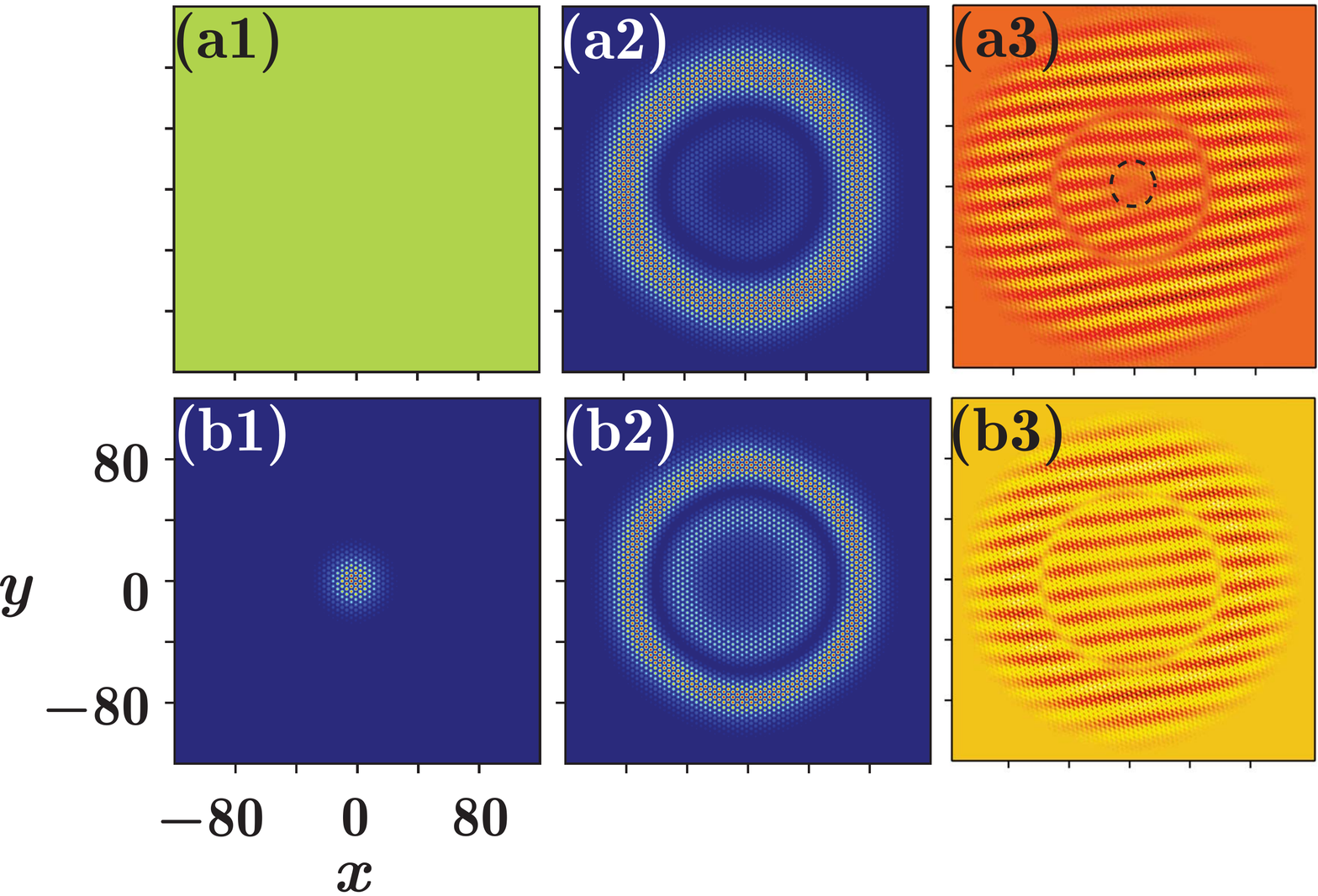}
  \caption{Same as Fig. \ref{fig5}, but with the sublattice $B$ excited.}
  \label{fig6}
\end{figure}

On the other hand, if sublattice \textit{B} is excited,
the situation will be in the opposite,
so that the pseudospin is converted into a vortex angular momentum on sublattice \textit{A},
with a topological charge $-1$,
as shown in Fig. \ref{fig6}.
As mentioned before, the pseudospin matrix for the Dirac cone at \textbf{K} in the super-honeycomb lattice is nontrivial,
and the projections on sublattices \textit{C}, \textit{D} and \textit{E} are not clear.
As a result, we only display the vortex generation on sublattices \textit{A} and \textit{B},
since the eigenmodes on them are separated.

Let us turn now to the pseudospin-1 Dirac cone at $\bf \Gamma$.
According to the matrices in Eqs. (\ref{eq8})-(\ref{eq10}) and the corresponding states in Eqs. (\ref{eq11a})-(\ref{eq11e}),
we have to consider two cases: the psedudospin-0 case and the psedudospin-1 case, even though we call it the pseudospin-1 Dirac cone.
Considering that the pseudospin connected with the Dirac cones at $\bf K$ and ${\bf K}'$ is $\pm1/2$,
which reveals the fermionic nature of the super-honeycomb lattice,
the pseudospin connected with the Dirac cone at $\bf \Gamma$ reflects its bosonic nature, because of the integer pseudospins 0 and $\pm1$.
Therefore, the super-honeycomb lattice exhibits both fermoinic and bosonic properties,
which is a consequence of its hybrid, honeycomb and Lieb, nature.

Considering that $\hat S _z$ is not diagonal in the natural sublattices,
we have to take the method used in the previous investigation \cite{diebel.prl.116.183902.2016} to check the pseudospin-mediated vortices.
We should note that the pseudospin-1 Dirac cone is located in the middle of the first Brillouin zone ($\bf \Gamma$),
which comes from the Lieb lattice \cite{leykam.pra.86.031805.2012,diebel.prl.116.183902.2016}.
So, there is no global phase tilt in our investigation,
and the output beam intensities corresponding to different pseudospins can be directly obtained from the output beam according to Eq. (\ref{eq11}).
Also, considering that both eigenstates $|S_z=0\rangle$ in Eq. (\ref{eq11c}) and $|S_z=\pm1\rangle$ in Eqs. (\ref{eq11d}) and (\ref{eq11e})
demand excitation of \textit{C}, \textit{D} and \textit{E}, which looks like they are somehow ``in entanglement'',
we only take the eigenstate $|S_z=0\rangle$ in Eq. (\ref{eq11a}) or (\ref{eq11b}) as an input.

\begin{figure*}[htpb]
\centering
 \includegraphics[width=\textwidth]{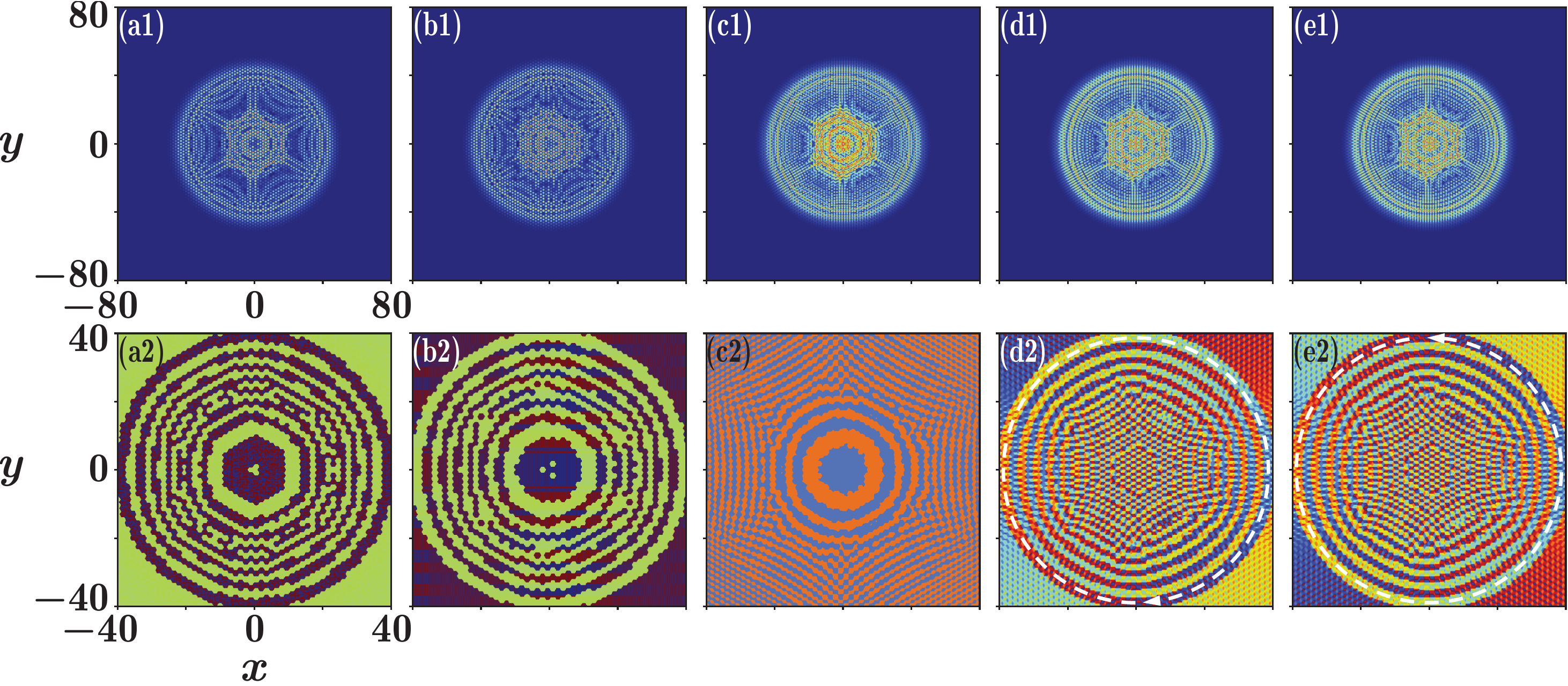}
  \caption{Pseudospin-mediated vortex generation when sublattice \textit{A} is excited according to Eq. (\ref{eq11a}).
  (a1)-(e1) Square root of the amplitudes of the output pseudospin components obtained according to Eqs. (\ref{eq11a})-(\ref{eq11e}), respectively.
  (a2)-(e2) Corresponding phase profiles.
  The dashed circle with an arrow indicates the phase gradient.}
  \label{fig7}
\end{figure*}

The results are displayed in Fig. \ref{fig7},
in which the panels in the first row are the square root of the amplitude of different components obtained according to Eqs. (\ref{eq11a})-(\ref{eq11e})
and the panels in the second row are the corresponding phase profiles.
The beams are not hollow conical beams, because they carry contributions from other modes of the bands,
but one may observe that they consist of many concentric rings.
These rings come just from the pseudospin components we are interested in, which indeed are the pseudospin mediated vortices.
To avoid checking the vortex and anti-vortex pairs in the complicated phase profiles,
the total topological charge can be obtained directly from the phase gradients of the rings.
In Figs. \ref{fig7}(a2)-\ref{fig7}(c2), there is no phase gradient in the rings,
so the topological charge is 0, and these rings are not vortices.
In Fig. \ref{fig7}(d2), there is a phase gradient from $\pi$ to $-\pi$, as indicated by the dashed circle with an arrow.
Therefore, the rings in Fig. \ref{fig7}(d1) form a vortex with a topological charge +1.
The phase gradient in Fig. \ref{fig7}(e2) is opposite that in Fig. \ref{fig7}(d2),
so the rings in Fig. \ref{fig7}(e1) form a vortex with a topological charge $-1$.
The results agree with those reported for a Lieb lattice \cite{diebel.prl.116.183902.2016}.

\section{Conclusion}
\label{conclusion}
In summary, we have introduced the super-honeycomb lattice,
which can be viewed as a combination of the honeycomb lattice and the Lieb lattice.
We have investigated transport properties in this lattice.
We identified a flat band, and pseudospin-1/2 and pseudospin-1 Dirac cones in the band structure of the super-honeycomb lattice.
Strong localization of light due to the flat band is observed,
and conical diffraction coming from the pseudospin-1/2 and pseudospin-1 Dirac cones is displayed.
We have also discussed the pseudospin-mediated vortex generation based on the pseudospin-1/2 and pseudospin-1 Dirac cones.
The super-honeycomb lattice is a hybrid fermionic and bosonic system,
which is also reported in \cite{lan.prb.85.155451.2012}.
This fact provides a new platform for investigating light trapping, higher pseudospin states, vortex generation,
and other interesting phenomena in this novel physical system.

We believe that other novel topological properties of the super-honeycomb lattice are ready for further exploration,
and deeper investigation on this interesting system may inspire new ideas and bring about new physical phenomena.

\section*{Acknowledgement}
The work was supported by the National Basic Research Program of China (2012CB921804),
National Natural Science Foundation of China (61308015, 11474228),
Key Scientific and Technological Innovation Team of Shaanxi Province (2014KCT-10),
and Qatar National Research Fund  (NPRP 6-021-1-005).
MRB also acknowledges support by the Al Sraiya Holding Group.

\bibliography{my_refs_library}
\bibliographystyle{myprx}

\end{document}